\documentclass[manuscript]{emulateapj}

\newcommand{\nature}{Nature}                       % Nature
\newcommand{\asr}{Adv.~Space~Res.}            % Advances in Space Research
\newcommand{\pss}{Planet.~Space~Sci.}         % Planetary & Space Science
\newcommand{\expa}{Exp.~Astron.}                % Experimental Astronomy

\shorttitle{The puzzling chemical composition of GJ~436b's atmosphere}
\shortauthors{Ag\'undez et al.}

\begin{document}

\title{The puzzling chemical composition of GJ~436b's atmosphere: \\
influence of tidal heating on the chemistry}

\author{Marcelino Ag\'undez\altaffilmark{1,2}, Olivia Venot\altaffilmark{3}, Franck Selsis\altaffilmark{1,2}, and Nicolas Iro\altaffilmark{4}}

\altaffiltext{1}{Univ. Bordeaux, LAB, UMR 5804, F-33270, Floirac, France; Marcelino.Agundez@obs.u-bordeaux1.fr}

\altaffiltext{2}{CNRS, LAB, UMR 5804, F-33270, Floirac, France}

\altaffiltext{3}{Instituut voor Sterrenkunde, KU Leuven, Celestijnenlaan 200D, 3001 Leuven, Belgium}

\altaffiltext{4}{Theoretical Meteorology group, Klimacampus, University of Hamburg, Grindelberg 5, 20144, Hamburg, Germany}

\begin{abstract}
The dissipation of the tidal energy deposited on eccentric planets may induce a heating of the planet that affects its atmospheric thermal structure. Here we study the influence of tidal heating on the atmospheric composition of the eccentric ($e$ = 0.16) "hot Neptune" GJ~436b, for which inconclusive chemical abundances are retrieved from multiwavelength photometric observations carried out during primary transit and secondary eclipse. We build up a one-dimensional model of GJ~436b's atmosphere in the vertical direction and compute the pressure-temperature and molecular abundances profiles for various plausible internal temperatures of the planet (up to 560 K) and metallicities (from solar to 100 times solar), using a radiative-convective model and a chemical model which includes thermochemical kinetics, vertical mixing, and photochemistry. We find that the CO/CH$_4$ abundance ratio increases with metallicity and tidal heating, and ranges from 1/20 to 1000 within the ranges of metallicity and internal temperature explored. Water vapour locks most of the oxygen and reaches a very high abundance, whatever the metallicity and internal temperature of the planet. The CO$_2$/H$_2$O abundance ratio increases dramatically with metallicity, and takes values between 10$^{-5}$-10$^{-4}$ with solar elemental abundances and $\sim$0.1 for a metallicity 100 times solar. None of the atmospheric models based on solid physical and chemical grounds provide a fully satisfactory agreement with available observational data, although the comparison of calculated spectra and observations seem to point to models with a high metallicity and efficient tidal heating, in which high CO/CH$_4$ abundance ratios and warm temperatures in the dayside atmosphere are favoured.
\end{abstract}

\keywords{planetary systems --- planets and satellites: atmospheres --- planets and satellites: composition --- planets and satellites: individual (GJ~436b)}

\section{Introduction}

GJ~436b is probably the most interesting Neptune analogue found to date among the zoo of known exoplanets. First discovered around the nearby M dwarf star GJ 436 by \cite{but2004} through the radial velocity method, and later on observed to transit its host star by \cite{gil2007}, it has been extensively studied at visible and infrared wavelengths \citep{man2007,demi2007,demo2007,alo2008,cac2009,pon2009,bal2010,ste2010,bea2011,knu2011}. With a mass of 23 $M_{\oplus}$, somewhat above that of Neptune, and an orbital distance of just 0.03 AU, GJ~436b has been labelled as a ''hot Neptune''. Indeed, although GJ 436 is significantly cooler than solar-type stars, the short orbital distance makes the planet to be highly irradiated, resulting in a planetary effective temperature of about 700-800 K \citep{demi2007,demo2007,ste2010}. An interesting aspect of this particular "hot Neptune" is that, unlike most close-in planets which have circular orbits, GJ~436b possesses a significant eccentricity of $\sim$0.16 \citep{man2007,demi2007,demo2007,knu2011}.

Information on the chemical state of GJ~436b's atmosphere has been obtained through multiwavelength photometry during the primary transit and secondary eclipse, thanks to the relatively large transit depth and favourable planet-to-star emission contrast. \cite{ste2010} were able to built up a dayside emission spectrum based on Spitzer photometric observations of the secondary eclipse at six wavelengths from 3.6 to 24 $\mu$m. Their analysis, and the more recent ones by \cite{mad2011} and \cite{mos2013} based on the same observation data, indicate that the atmosphere of GJ~436b is rich in carbon monoxide ($>$1000 ppm), but poor in methane ($<$1 ppm) and water vapour ($<$100 ppm). In an independent study, however, \cite{bea2011} reported primary transit observations obtained with Spitzer at 3.6, 4.5, and 8 $\mu$m and found that the transmission spectrum is consistent with an atmosphere mainly composed of methane and molecular hydrogen. To further complicate the picture, recently \cite{knu2011} re-analysed the same set of primary transit observations reported by \cite{bea2011} and obtained significantly different transit depths, especially at 3.6 and 4.5 $\mu$m, which led \cite{knu2011} to favour an atmospheric composition with enhanced CO and reduced CH$_4$, consistent with the findings of \cite{ste2010} on the planet dayside. These contradictory conclusions suggest that it is certainly too soon to draw a definite picture on the chemical composition of GJ~436b's atmosphere. New observations, especially those involving future space missions such as the James Webb Space Telescope \citep{gar2006} and EChO \citep{tin2012}, are clearly needed.

In the mean time, theoretical models can also provide valuable insights on the atmospheric composition of GJ~436b. Recently, \cite{lin2011} developed a detailed model of the atmosphere of GJ~436b including thermochemical kinetics, vertical mixing, and photochemistry, and concluded that methane should be the major carbon-bearing molecule under most plausible conditions. Here we revisit the chemistry of GJ~436b's atmosphere. In particular, we investigate the effects that tidal heating caused by the eccentric orbit of the planet may have on the thermal structure of the atmosphere, and in turn on its chemical composition. Concurrent with this work, \cite{mos2013} have recently published an independent chemical study of GJ~436b's atmosphere in which metallicities up to 10,000 times over solar are explored.

\section{The atmosphere model} \label{sec:model}

\begin{table}
\caption{GJ~436b's parameters} \label{table:parameters}
\centering
\begin{tabular}{l@{\hspace{1.5cm}}r}
\hline \hline
Parameter                                                   & Value \\
\hline
Stellar radius                                              & 0.455 $\pm$ 0.018 $R_\odot$$^a$ \\
Stellar effective temperature                      & 3416 $\pm$ 54 K$^a$ \\
Planetary radius                                         & 4.09 $\pm$ 0.20 $R_{\oplus}$$^b$ \\
Planetary mass                                          & 23.4 $\pm$ 1.6 $M_{\oplus}$$^b$ \\
Orbital semimajor axis                              & 0.02887 $\pm$ 0.00089 AU$^b$ \\
%Stellar type                                                & M3V$^a$ \\
%gravity                                                       & 13.7 $\pm$ 1.1 m s$^{-2}$$^b$ \\
%Planetary radius                                         & 0.365 $\pm$ 0.018 $R_{\rm Jupiter}$$^b$ \\
%Planetary mass                                          & 0.0737 $\pm$ 0.0051 $M_{\rm Jupiter}$$^b$ \\
\hline
\end{tabular}
\tablerefs{$^a$ \citeauthor{von2012} (2012); $^b$ \citeauthor{sou2010} (2010).}
\end{table}

We model the atmosphere of GJ~436b as a one-dimensional column in the vertical direction, adopting the planetary and stellar parameters given in Table~\ref{table:parameters}. The model therefore neglects any possible variation of temperature and chemical composition with longitude and latitude. This assumption may need to be revised in the future given the non-uniform temperature distribution predicted for GJ~436b by the general circulation model (GCM) of \cite{lew2010}, especially if the metallicity of the atmosphere is substantially higher than solar, and the possibility that primary transit and secondary eclipse observations probe regions, limb and dayside respectively, which may have significant differences in their physical and chemical conditions (e.g. \citealt{agu2012,agu2013}).

\begin{figure}
\centering
\includegraphics[angle=0,width=\columnwidth]{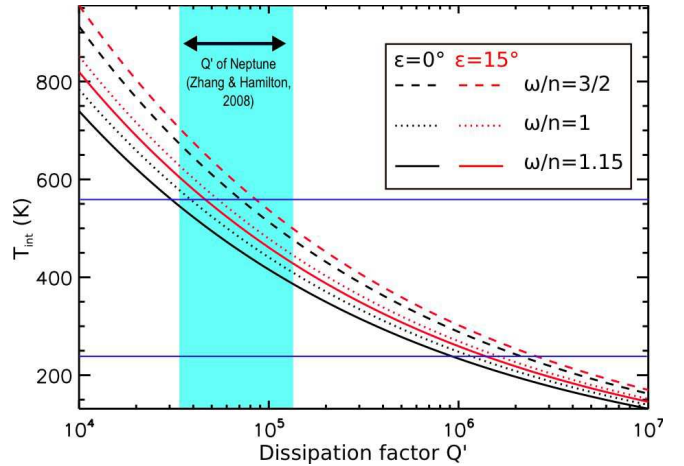}
\caption{Internal temperature as a function of the tidal dissipation factor $Q'$. Calculations were done for two obliquities (0 and $15^\circ$) and three rotation rates predicted by tidal models: two spin orbit resonances (1:1 and 3:2) and the pseudo-synchronization \citep{lev2007}. The two horizontal lines indicate the internal temperatures used in the model.} \label{fig:tidal-heating}
\end{figure}

A key parameter of the model, and a valuable information we want to retrieve from observations, is the elemental composition of the atmosphere, which is highly uncertain. There are arguments which permit to think of a significant enrichment in heavy elements with respect to the values of the host star, whose metallicity is nearly solar (\citealt{one2012} and references therein). Given the mass and radius of GJ~436b, its bulk composition must be enriched in elements heavier than hydrogen and helium according to models of the planet interior \citep{net2010}. The atmosphere of such a planet is also likely to be enhanced in heavy elements due to the reduced efficiency to retain light elements, compared with Jupiter-mass planets \citep{elk2008}. In the solar system, for example, the abundance of carbon is enhanced over the solar value by just a factor of 3 in Jupiter and Saturn and by a factor of about 50 in Uranus and Neptune \citep{her2004}. The metallicity in the atmosphere of "hot Neptunes" may be even higher taking into account that Neptune is believed to contain an ice rich mantle, a fraction of which could be vapourized in the case of "hot Neptunes" producing a further enrichment of heavy elements in the gaseous envelope of the planet. Given the above arguments, we consider three different elemental compositions for the atmosphere of GJ~436b, in which the abundance of elements heavier than helium are enhanced over the solar values compiled by \cite{asp2009} by factors of $\zeta$ = 1, 10, and 100. We therefore restrict our calculations to a solar elemental C/O abundance ratio.

Due to its eccentricity of 0.16 \citep{man2007} and its proximity to its host star, GJ~436b is subjected to strong tidal forces. The resulting rate of dissipation in the planet depends on the unknown internal composition and structure. In the \textit{Constant Phase Lag} (CPL) tidal model, dissipation is determined by the quantity $Q/k_2$, where $Q$ is the quality factor and $k_2$ is the Love number of degree 2 \citep{gol1966}. The two quantities $Q$ and $k_2$ are often merged into one single parameter, the reduced Q-value, $Q'=3Q/2k_2$. In the \textit{Constant Time Lag} (CTL) model, dissipation is controled by the quantity $k_2 \Delta t$, where $\Delta t$ is the time lag \citep{hut1981}. In this study we use the CTL model but, in order to make easier the comparison with previous studies dedicated to GJ~436b, we use $Q'$ as the dissipation factor, using the relation $k_2 \Delta t =3/(2Q'n)$, where $n$ is the orbital mean motion \citep{lec2010}. The dissipation rate $\dot{E}$ is given in the CTL approach by equation (A26) of \cite{lec2010}. The internal heat flow $\phi_{int}=\dot{E}/(4\pi R_p^2$), where $R_p$ is the planetary radius, released by tidal dissipation is introduced into the radiative-convective model (see below) as an internal temperature $T_{int}=(\phi_{\rm int}/\sigma)^{1/4}$, where $\sigma$ is the Stefan-Boltzmann constant. Fig.~\ref{fig:tidal-heating} shows the internal temperature as a function of the dissipation factor $Q'$ for different rotation rates, the 1:1 and 3:2 spin-orbit resonances and the pseudo-synchronization (a rotation rate predicted by tidal models, which minimizes the dissipation; see \citealt{lec2010}), and two obliquities, 0 and $15^\circ$. Due to its efficient tidal erosion a non zero obliquity would imply the influence of a planetary companion resulting in a Cassini state \citep{col1966,pea1969}. Values of $Q'$ for Neptune have been constrained between $3.3\times10^4$ and $1.35\times10^5$, based on the evolution of its system of satellites \citep{zha2008}. However, the dissipation factor can vary by orders of magnitude  from one planet to another and constraining the actual dissipation rate in GJ~436b is challenging. One way could be to put an upper limit on the intrinsic emission of the planet \citep{demi2007} but establishing a robust radiative budget of the planet would require a comprehensive survey in terms of wavelengths and geometries. We will indeed see in section~\ref{sec:emission} that the photosphere of the planet in the spectral bands of available observations is located significantly above the layers affected by the internal heat flux. Another way is to compare the age of the system (1-10 Gyrs, Torres et al., 2007) with the circularisation timescale $\tau_e=e/\dot{e}$. In good agreement with \citet{mar2008}, we find that only values of $Q'$ above $10^5$ are compatible with the observed eccentricity, when considering GJ~436b as the only planet of its host star. With $Q'=10^5$ we find an internal temperature $T_{int} < 560$~K. As suggested by \cite{beu2012}, Kozai migration due to the interactions with a companion could have considerably slowed down the decrease of the eccentricity of planet b. An even lower value of $Q'$ and thus a stronger tidal heating are therefore possible, although we conservatively use a maximum internal temperature of 560~K. For our modeling, we consider internal temperatures of 240, 400, and 560 K, in the range of the values expected according to the above arguments, as well as a low value of 100 K, which is adopted here just for comparison purposes.

In order to explore the sensitivity of the atmospheric chemical composition to the metallicity and efficiency of tidal heating we have considered a total number of 3 $\times$ 4 cases in which $\zeta$ = 1, 10, and 100, and $T_{int}$ = 100, 240, 400, and 560 K. The one-dimensional vertical model of GJ~436b's atmosphere is divided into two parts. The first one consists of a radiative-convective model, in which a vertical temperature profile is calculated. The second one is a chemical model which uses as input the previously calculated temperature profile, and which includes thermochemical kinetics, vertical diffusion, and photochemistry, providing as output the vertical abundance profiles of the different species considered.

The radiative-convective model has been already described \citep{iro2005,iro2010,agu2012}. The temperature profile in the vertical direction is computed at radiative equilibrium between 10$^3$ and 10$^{-6}$ bar (for upper layers we consider an isothermal atmosphere). We adopt a mean insolation over the planet's orbit, i.e. we neglect any variation of the incoming stellar flux during an orbital period due to the non-zero eccentricity of GJ~436b. This effect has been studied by \cite{iro2010} for various eccentric planets. Based on these calculations we expect temperature variations of less than 100 K during the orbital period for a moderately eccentric planet ($e$ = 0.16) such as GJ~436b. The incident stellar flux is calculated from a Phoenix synthetic spectrum \citep{hau1999}\footnote{See \texttt{http://phoenix.ens-lyon.fr/Grids/NextGen/.}} for a star with an effective temperature of 3400 K, a surface gravity of 10$^{5.0}$ cm s$^{-2}$, and solar metallicity, and adopting the semimajor axis given in Table~\ref{table:parameters} as the mean planet-star distance. We assume complete redistribution of the incident stellar flux over the whole planet. According to the GCM simulations of \cite{lew2010}, heat is efficiently redistributed from the day to the night side if the atmosphere has a solar metallicity, although the redistribution turns less efficient if the metallicity is increased 50 times over solar. Since the metallicity of GJ~436b's atmosphere is highly uncertain, it is not clear whether it is more appropiate to choose a redistribution over the whole planet or over just the dayside hemisphere. In their one-dimensional radiative-convective models, \cite{lew2010} use the former choice, while the latter is adopted by \cite{mos2013}. The sources of atmospheric opacity included are Rayleigh scattering, collision-induced absorption from H$_2$--H$_2$ and H$_2$--He pairs, bound-free absorption from H$^-$, free-free absorption from H$_2^-$, and spectroscopic lines of CO, H$_2$O, CH$_4$, CO$_2$, NH$_3$, TiO, and the alkali atoms Na and K. In a first step, the abundances of the absorbing species are calculated at thermochemical equilibrium. The resulting temperature profile is then used by the chemical model (see below) to compute non-equilibrium abundances for CO, H$_2$O, CH$_4$, CO$_2$, and NH$_3$ (the rest of absorbing species are not included in the chemical model), which are then used in a second iteration of the radiative-convective model to recompute the temperature profile. Corrections to the temperature due to the use of non-equilibrium, instead of chemical equilibrium, abundances are moderate ($<$100 K), mainly because H$_2$O, the main species affecting the thermal structure, has an abundance close to chemical equilibrium (see Section~\ref{sec:abundances}). The extent and sign of the corrections depend on the specific parameters ($\zeta$ and $T_{int}$) of each model and on the pressure level, although a general conclusion is that corrections become more important as metallicity increases.

\begin{figure}
\centering
\includegraphics[angle=0,width=\columnwidth]{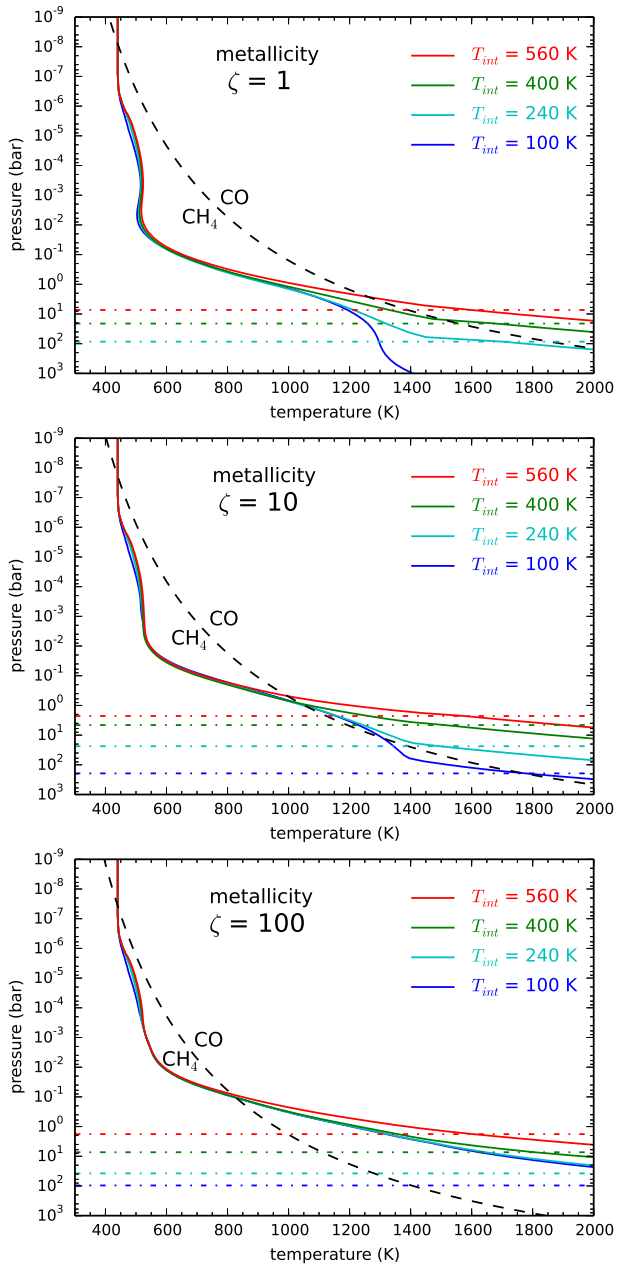}
\caption{Vertical temperature profile calculated for GJ~436b's atmosphere for three different metallicities: $\zeta$ = 1, 10, and 100 (upper, middle, and lower panels, respectively), and for various internal temperatures $T_{int}$ of the planet. The horizontal dot-dashed lines indicate the transition between the convective and radiative parts of the atmosphere. Black dashed lines delimitate the two regions where either CH$_4$ or CO is the major carbon reservoir at chemical equilibrium. Note that an increase in $T_{int}$ and also in $\zeta$ makes the thermal profile (especially in the bottom layers) to move into the region dominated by CO.} \label{fig:tk}
\end{figure}

The resulting temperature profiles for the 3 $\times$ 4 cases studied are shown in Fig.~\ref{fig:tk}. In the upper atmospheric layers, above the 10$^{-2}$ bar pressure level, temperatures are not very different among the different cases investigated, and are consistent within 100 K with those calculated by \cite{lew2010} in absence of dynamics and under the same hypothesis adopted by us of planet-wide redistribution of incident energy. The atmosphere is radiative over most of its vertical structure, although in the deep layers a convective region appears and the temperature gradient becomes adiabatic. The convective-to-radiative transition occurs at pressures between $>$1 kbar and 2 bar, depending on the metallicity and internal temperature (see Fig.~\ref{fig:tk}). An increase in the internal temperature of the planet as a consequence of an efficient tidal heating makes the hot convective region to shift upwards in the atmosphere, and overall results in higher temperatures in the bottom of the atmosphere. An enhanced metallicity plays a somewhat similar role, shifting the convective region to lower pressures and increasing the temperatures in the deep layers, an effect that has been also found by \cite{lew2010} for GJ~436b. A major effect on an increase in both the internal temperature of the planet and the metallicity is that the thermal profile moves into the region where carbon is mostly as CO instead of as CH$_4$ at thermochemical equilibrium.

The chemical model, which has been previously applied to study the atmosphere of "hot Jupiters" \citep{ven2012}, considers a vertical column of atmosphere and solves the equation of continuity as a function of time until a steady state is reached. It is therefore a one-dimensional model in which thermochemical kinetics, vertical transport, and photochemistry are taken into account. The chemical network has been validated in the area of combustion chemistry and includes about 100 neutral species linked by about 2000 chemical reactions (see details in \citealt{ven2012}). The temperature-pressure profile is calculated via the radiative-convective code, as described above, for each of the 3 $\times$ 4 cases under study. Since the temperatures at depths corresponding to 1 kbar are well above the validity range of our chemical network (300-2500 K) in many of the 3 $\times$ 4 cases covered, in the chemical model we set the bottom of the atmosphere to a level where the temperature does not significantly exceed 2500 K. The location of the lower boundary is of not importance for the calculated abundances provided it remains well below the quench level.

As usual, vertical transport is modeled as a diffusive process characterized by an eddy diffusion coefficient, which in the case of exoplanet atmospheres is solely constrained by GCM models. Based on their GCM model of GJ~436b, \cite{lew2010} estimate an eddy diffusion coefficient of 10$^8$ cm$^2$ s$^{-1}$ at 100 bar and of 10$^{11}$ cm$^2$ s$^{-1}$ at 10$^{-4}$ bar, by multiplying a mean vertical wind speed by the local scale height. In the absence of better constraints, we adopt these latter values and assume a log-log linear behaviour of the eddy diffusion coefficient with pressure in the 100-10$^{-4}$ bar range, and a constant value at $<$10$^{-4}$ bar. In the convective region of the atmosphere, whose exact location depends on the metallicity and internal temperature of the planet (see Fig.~\ref{fig:tk}), we impose a relatively high value of 10$^{10}$ cm$^2$ s$^{-1}$. The eddy diffusion coefficient, although poorly constrained, is a key parameter as it determines the quench level where abundances depart from chemical equilibrium, and therefore the values at which molecular abundances remain quenched along much of the atmosphere. For simplicity, here we do not explore the sensitivity of the chemical composition to the eddy diffusion coefficient, although it is wise to keep in mind these considerations when interpreting the molecular abundances resulting from the chemical model.

Photochemistry is also an important process in the upper dayside atmosphere. The photodissociation cross sections adopted here are described in \citeauthor{ven2012} (2012). The ultraviolet spectrum of GJ 436 has been recently observed in the 115-310 nm wavelength range with the Hubble Space Telescope \citep{fra2013}, and is adopted here. At wavelengths shortward of 115 nm we adopt the Sun spectrum (mean between maximum and minimum activity; \citealt{thu2004}), while at wavelengths longward of 310 nm we take the same Phoenix synthetic spectrum used in the radiative-convective model.

\section{Calculated molecular abundances} \label{sec:abundances}

\begin{figure*}
\centering
\includegraphics[angle=0,width=\textwidth]{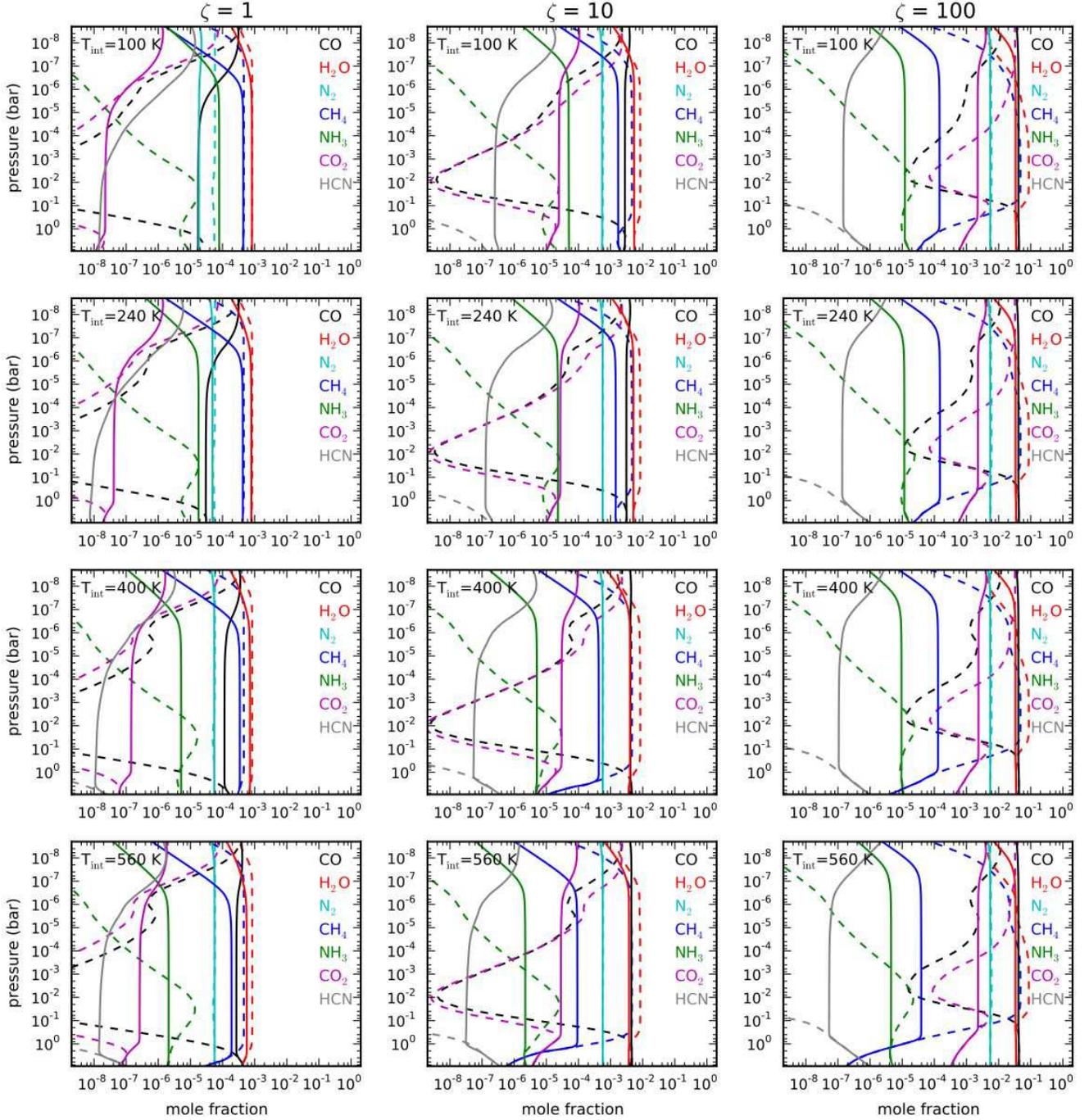}
\caption{Vertical distribution of molecular abundances as calculated with the chemical model that includes thermochemical kinetics, vertical mixing, and photochemistry (solid lines) and at chemical equilibrium (dashed lines). The different panels correspond to different choices of metallicity $\zeta$ = 1, 10, and 100 (panels at the left, centre, and right, respectively) and internal temperature of the planet $T_{int}$ = 100, 240, 400, and 560 K (panels from top to bottom).} \label{fig:abundances}
\end{figure*}

In Fig.~\ref{fig:abundances} we compare the calculated vertical abundance profiles of some of the most abundant molecules with their chemical equilibrium vertical profiles, for the different cases of metallicity ($\zeta$ = 1, 10, and 100) and internal temperature of the planet ($T_{int}$ = 100, 240, 400, and 560 K) investigated. 

A first important prediction is that water vapour is present with a very high abundance, whatever the adopted metallicity and internal temperature of the planet. Basically, H$_2$O locks nearly all available oxygen or, at worst, the excess of oxygen not locked into CO (i.e. the excess of oxygen over carbon). This result is in agreement with the findings of \cite{lin2011} and \cite{mos2013}, who also predict very high H$_2$O abundances in GJ~436b's atmosphere whatever the metallicity. The high abundance predicted for H$_2$O along most of the atmosphere, anywhere below the photochemical active region, is mostly based on thermochemical grounds and is therefore insensitive to the location of the quench level and consequently to the choice of the eddy diffusion coefficient profile. Unless the atmosphere of GJ~436b has an elemental C/O abundance ratio above unity or a metallicity in excess of $\zeta$ = 100, chemical models predict that water vapour is nearly the most abundant species after H$_2$ and He, and therefore the major atmospheric absorber at infrared wavelengths.

The chemical quantity that is most influenced by the metallicity and internal temperature of the planet is probably the CO/CH$_4$ abundance ratio. It is essentially determined by the chemical equilibrium abundances of CO and CH$_4$ around the quench level, which is located somewhere between 1 and 10 bar for our choice of eddy diffusion coefficient in most of the cases. An increase in the internal temperature of the planet makes the bottom layers to be hotter so that they penetrate deeper into the region where CO dominates over CH$_4$ at chemical equilibrium (see Fig.~\ref{fig:tk}). As a consequence the abundance at which CO gets quenched increases at the expense of CH$_4$. For example, under solar elemental abundances ($\zeta$ = 1) CH$_4$ is more abundant than CO if $T_{int}$ = 100 K, but if $T_{int}$ = 560 K then it is CO which dominates (see the trend of black and blue solid lines with increasing $T_{int}$ in Fig.~\ref{fig:abundances}). An enhancement in the metallicity also favours CO over CH$_4$, on the one side because it also results in a warming of the deep atmosphere, and on the other because it shifts the CO:CH$_4$ chemical equilibrium transition to lower temperatures (e.g. the temperature at which CO and CH$_4$ have equal abundances at 1 bar is lowered from $\sim$1160 K to $\sim$960 K when the metallicity is enhanced from $\zeta$ = 1 to $\zeta$ = 100; see Fig.~\ref{fig:tk}). In the atmosphere of GJ~436b, the CO/CH$_4$ abundance ratio is therefore extremely sensitive to the metallicity and also to the efficiency of tidal heating. In the cases explored the CO/CH$_4$ abundance ratio spans over a range of values between 0.05 (at solar metallicity and low $T_{int}$) and 1000 (at $\zeta$ = 100 and high internal temperature). This range is likely to be even wider due to the uncertainty associated to the eddy diffusion coefficient, which is key in determining the level at which the abundances of CO and CH$_4$ get quenched.

Our results concerning the CO/CH$_4$ abundance ratio are slightly different from those presented by \cite{lin2011} and \cite{mos2013}. In their model with solar metallicity, the calculated abundances are of the same order than those found by us in the case of $\zeta$ = 1 and $T_{int}$ = 100 K, except for a lower mixing ratio for CO in their model (around 10-100 times less than our calculated value). The discrepancy is likely due to differences in the temperatures below the 1 bar pressure level (temperatures come from the GCM by \citealt{lew2010} in their case), although differences in the adopted eddy diffusion coefficient and in the chemical network can also contribute.

The system N$_2$/NH$_3$ behaves in a similar fashion to the CO/CH$_4$ one, with molecular nitrogen being favoured at high temperatures and ammonia at low temperatures. In the cooler model (that with $\zeta$ = 1 and $T_{int}$ = 100 K) NH$_3$ is more abundant than N$_2$, although as the metallicity increases and tidal heating becomes more efficient N$_2$ becomes the major nitrogen carrier at the expense of ammonia. The N$_2$/NH$_3$ abundance ratio takes values from 0.2, at solar metallicity and low internal temperature, up to 1000, at $\zeta$ = 100 and $T_{int}$ =560 K. However, similarly to the case of CO/CH$_4$, the range is likely to be wider due to the sensitivity of the quench level of NH$_3$ to the eddy diffusion coefficient, which is not well constrained.

Carbon dioxide is also an abundant molecule which has quite characteristic spectral signatures at infrared wavelengths. The abundance of CO$_2$ has just a mild dependence with the internal temperature of the planet (see Fig.~\ref{fig:abundances}), although it is extremely sensitive to the metallicity, as has been already recognized for "hot Jupiters" and "hot Neptunes" \citep{zah2009,lin2011,mos2013}. Carbon dioxide contains three heavy atoms and thus its abundance scales as $\zeta^2$ or even as $\zeta^3$, depending on the exact thermal structure of the atmosphere.

The rate of vertical transport being relatively high due to our choice of eddy diffusion coefficient, much of the atmosphere is characterized by vertically flat abundance profiles, whose values are determined at the quench level (located between 1 and 100 bar depending on the molecule). The layers where photochemistry is active are confined to pressures below 10$^{-6}$ bar, with the main effects being the destruction of NH$_3$, CH$_4$, and H$_2$O, and the stimulated formation of HCN, CO$_2$, and CO. These effects are similar to those found by \cite{lin2011}, although the extent of the photochemically active region and of the photochemical effects themselves is much more important in their models, probably because they adopt a much smaller eddy diffusion coefficient in the upper layers (10$^8$ cm$^2$ s$^{-1}$) than us (10$^{11}$ cm$^2$ s$ ^{-1}$). This is also likely the reason of the much more important effect of molecular diffusion in their models, where all species heavier than atomic hydrogen are severely depleted above the 10$^{-7}$ bar level, than in ours, where molecular diffusion is not even at work at about 10$^{-9}$ bar. In support of this explanation, we note that the chemical model of GJ~436b's atmosphere by \cite{ven2013a}, which also uses an eddy diffusion coefficient of 10$^8$ cm$^2$ s$^{-1}$ throughout the whole atmosphere, finds the same extent of photochemistry and molecular diffusion than the models by \cite{lin2011}. In the study by \cite{mos2013}, their choice of an eddy diffusion coefficient of 10$^9$ cm$^2$ s$^{-1}$ makes the extent of photochemistry to be intermediate between the models of \cite{lin2011} and ours.

Whether photochemistry has a larger extent than in the models presented here, it is unlikely to be due to a bad choice of the stellar ultraviolet spectrum, which is well constrained from observations of GJ~436 \citep{fra2013}, although it may well be due to an uncorrect choice of the eddy diffusion coefficient in the upper layers. To this regard, we note that \cite{par2013} have recently estimated an effective eddy diffusion coefficient by means of passive tracers in a GCM of the "hot Jupiter" HD 209458b and have found values which are about 10-100 times lower than those previously estimated through the more crude recipe of multiplying a mean vertical wind speed by the local scale height. It could therefore be plausible that the eddy diffusion coefficient adopted by us for GJ~436b would need to be revised downward, in which case photochemistry would certainly have a larger extent than in the models presented here.

\section{Planetary spectra} \label{sec:spectra}

\begin{figure*}
\centering
\includegraphics[angle=0,width=\textwidth]{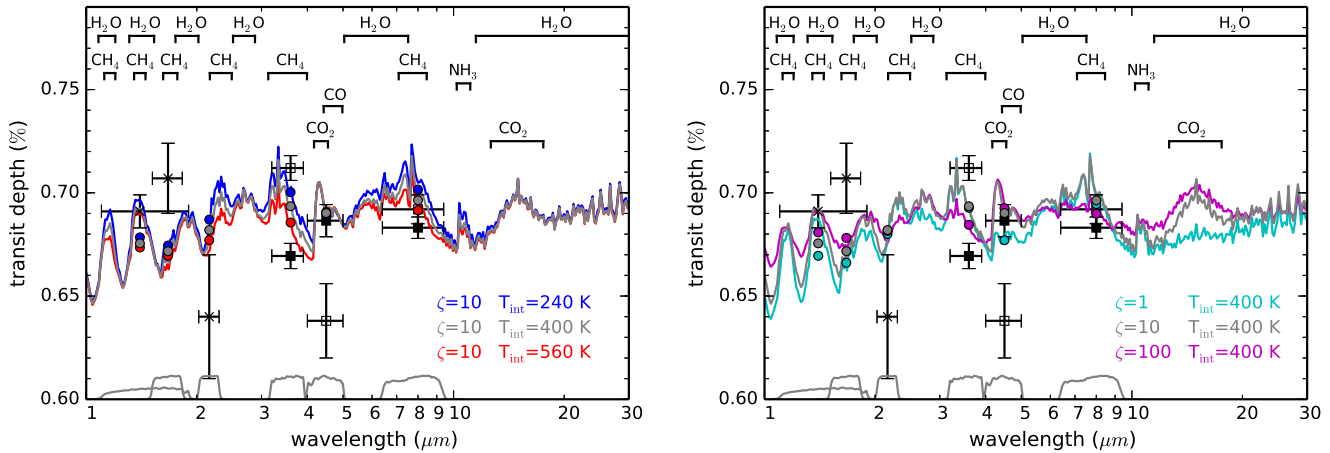}
\caption{Transmission spectra calculated for GJ~436b showing the effect of the internal temperature of the planet for $\zeta$ = 10 (left panel) and of the metallicity for $T_{int}$ = 400 K (right panel). The transit depth is simply calculated as $(R_p (\lambda) / R_*)^2$, where $R_p (\lambda)$ is the calculated radius of the planet as a function of wavelength and $R_*$ is the stellar radius, whose value is taken from Table~\ref{table:parameters}. Spectra are binned to a resolving power of $R$ = 100. The positions of the main absorption bands of H$_2$O, CH$_4$, CO, CO$_2$, and NH$_3$ are indicated. Photometric observations by HST NICMOS in the 1.1-1.9 $\mu$m range \citep{pon2009}, ground-based $H$ band \citep{alo2008}, and $K_s$ band \citep{cac2009} are shown as crosses, and Spitzer-IRAC photometric observations at 3.6, 4.5, and 8 $\mu$m from two different studies are shown as empty squares \citep{bea2011} and filled squares \citep{knu2011}. Grey lines at the bottom show the response of the filters used in the photometric observations. Calculated transit depths averaged over the various filters are shown as filled circles.} \label{fig:transmission}
\end{figure*}

There are significant constraints on the thermal and chemical structure of GJ~436b's atmosphere coming from multiwavelength photometric observations obtained during primary transit and secondary eclipse \citep{alo2008,cac2009,pon2009,ste2010,mad2011,bea2011,knu2011}. The interpretation of these data is however controversial. For example, the analysis of primary transit data carried out by \cite{bea2011} points toward an atmosphere rich in CH$_4$ and poor in CO while a re-analysis of the same data by \cite{knu2011} indicates the contrary. The retrieval of abundances from these photometric observations ends up with a set of molecular abundances which lack consistency among them from a chemical point of view. A coherent collection of molecular abundances can be obtained from a chemical model of the atmosphere, although it is by no means warranted that it may result in planetary spectra compatible with the observations. Unfortunately, \cite{lin2011} did not explore in their study how would the planetary spectra look like according to their calculated molecular abundances, so that the comparison with the observational studies had to rely on a comparison between retrieved and calculated abundances for individual molecules. The issue has been addressed in the recent study by \cite{mos2013}.

In order to investigate the degree of agreement between the predictions of the chemical model and the observation data, here we have computed the transmission and emission spectra of the planet adopting the one-dimensional vertical profiles of temperature and abundances resulting from the radiative-convective and chemical models. We have developed a line-by-line radiative transfer code in which the atmosphere is divided in several layers in the vertical direction (typically 60) and each layer is assumed to be homogeneous with longitude and latitude, so that it is characterized by a given pressure, temperature, and chemical composition. The absorption coefficient is calculated as a function of wavelength for each layer. Then, we calculate the transmission spectrum by computing, as a function of wavelength, the radius at which the tangential optical depth across the atmosphere equals 2/3. The emission spectrum of the planet is calculated by solving the equation of radiative transfer and averaging the emission over the planetary disk. The code has been checked against the suite of radiative transfer tools 'kspectrum'\footnote{See \texttt{http://code.google.com/p/kspectrum/}}, which has been widely used to model the atmosphere of Solar System planets such as Venus \citep{eym2009}. The sources of opacity included in the model are collision induced absorption by H$_2$--H$_2$ \citep{bor2001,bor2002} and H$_2$--He \citep{bor1989,bor1997,borfro1989}, and spectroscopic transitions of H$_2$O, CO, and CO$_2$ (from HITEMP; \citealt{rot2010}), and CH$_4$, NH$_3$, and HCN (from HITRAN; \citealt{rot2009}). Light scattering, which becomes important at wavelengths shorter than 1 $\mu$m, is not taken into account in the model. A more detailed description of the code is given in \cite{agu2013}.

\subsection{Transmission spectra}

In Fig.~\ref{fig:transmission} we show the transmission spectra calculated for various cases and compare them with available photometric observations of GJ~436b during primary transit conditions. We first note that setting the planetary radius of 4.09 $R_{\oplus}$ (see Table~\ref{table:parameters}) at the 1 bar level results in transit depths substantially higher than observed. We have therefore set the radius of 4.09 $R_{\oplus}$ at the 10 mbar level, in order to bring to a closer agreement the calculated spectra and the observed transit depths. This is an arbitrary choice that just serves to establish the absolute scale of the calculated transmission spectra and does not affect the relative variation of the transit depth with wavelength.

The left panel in Fig.~\ref{fig:transmission} shows the effect of the internal temperature of the planet for an intermediate choice of metallicity ($\zeta$ = 10). The main effect of an enhancement of tidal heating on the transmission spectrum is related to the depletion of the hydrides CH$_4$ and NH$_3$ (see Fig.~\ref{fig:abundances}), which makes the transit depth to decrease at the specific wavelengths at which these molecules absorb, around 1.1, 2,3, 3.3, and 7.7 $\mu$m in the case of CH$_4$ and at 10 $\mu$m for NH$_3$. In the right panel of Fig.~\ref{fig:transmission} we show the effect of metallicity on the transmission spectrum for an intermediate value of internal temperature ($T_{int}$ = 400 K). An increase in the metallicity causes an overall flattening of the transmission spectrum due to the increase of the mean mass of particles in the atmosphere (2.32, 2.50, and 4.14 Da for $\zeta$ = 1, 10, and 100, respectively), which makes the atmospheric scale height to be reduced. This flattening effect is also found in the case of the transmission spectra calculated by \cite{ven2013b} for GJ~3470b, and in the study of GJ~436b by \cite{mos2013}, where metallicities above 100 times solar are also explored. Appart from this change in the overall spectral shape, metallicity has also some effects on the transmission spectrum due to purely chemical reasons, that is, due to the relative variations in the abundances of the main molecules that provide atmospheric opacity. An increase of the metallicity from $\zeta$ = 1 to 100 produces a significant increase in the CO/CH$_4$ and N$_2$/NH$_3$ abundance ratios (see Fig.~\ref{fig:abundances}), which translates to enhanced CO absorption around 4.7 $\mu$m and depressed absorption by CH$_4$ at 2,3, 3.3, and 7.7 $\mu$m and by NH$_3$ around 10 $\mu$m. Increasing the metallicity brings also a spectacular increase of the CO$_2$/H$_2$O abundance ratio (see Fig.~\ref{fig:abundances}), leaving a clear signature in the transmission spectrum as an enhancement of the transit depth around 4.3 and 15 $\mu$m.

As shown in Fig.~\ref{fig:transmission}, the agreement between calculated and observed transit depths is somewhat poor. At wavelengths shorter than 3 $\mu$m, none of the models is able to account for the relative variation of the transit depths observed in the HST/NICMOS, $H$, and $K_s$ bands. For example, the absorption in the $K_s$ band at 2.16 $\mu$m is predicted slightly higher than in the $H$ band at 1.67 $\mu$m, while observations indicate the contrary. The error bars of these observations are however substantial and may limit their ability to put significant constraints on the chemical composition of GJ~436b's atmosphere. Even if their uncertainties were significantly improved, observations at these short wavelengths can barely distinguish between the different scenarios of internal temperature and metallicity, except perhaps for the important flattening of the spectrum seen at high metallicities. A similar behaviour is found in the high metallicity models of \cite{mos2013} regarding these short wavelength bands. Moreover, forward models in which temperature and chemical composition are varied to find the best match with observations \citep{knu2011,sha2011} find also a poor agreement with the primary transit observations in the HST/NICMOS, $H$, and $K_s$ bands.

The major constraints on the chemical composition of GJ~436b's atmosphere obtained from primary transit observations come probably from the Spitzer-IRAC transit depths at 3.6 and 4.5 $\mu$m. There are however large differences between the values from \cite{bea2011} and \cite{knu2011}, which do not allow to conclude about the most likely chemical composition. \cite{bea2011} find a higher transit depth at 3.6 $\mu$m than at 4.5 $\mu$m, which points to an atmosphere rich in methane and poor in carbon monoxide and carbon dioxide. This scenario is favoured in the models with low metallicity and low tidal heating, in which the CO/CH$_4$ and CO$_2$/H$_2$O abundance ratios are the lowest, although none of the models is able to reproduce the large difference in the transit depths observed by \cite{bea2011}. On the other hand, \cite{knu2011} find a slightly higher absorption at 4.5 $\mu$m than at 3.6 $\mu$m, which favours a CH$_4$--poor atmosphere. Those models in which tidal heating and metallicity are high would be consistent with a smaller transit depth at 3.6 $\mu$m than at 4.5 $\mu$m, as found by \cite{knu2011}. Unless more accurate primary transit spectra are obtained, it is difficult to draw a definitive conclusion about which are the major atmospheric constituents of GJ~436b based on the transmission spectrum. To this purpose, observations with the Hubble Space Telescope and future facilities such as the James Webb Space Telescope and EChO will be of great help \citep{sha2011}.

\subsection{Emission spectra} \label{sec:emission}

If we assume zero albedo and redistribution of the stellar energy absorbed over the entire planet, the equilibrium temperature $T_{eq}$ of GJ~436b is 650 K at an orbital distance equal to the semimajor axis. Because the planet possesses also an internal temperature independent of the stellar flux captured, the bolometric flux emitted by the planet is characterized by an effective temperature given by ($T_{eq}^4$ + $T_{int}^4$)$^{1/4}$, i.e. in the range 650-720 K for values of $T_{int}$ below 560 K. Therefore, tidal heating would produce a moderate increase of the planetary effective temperature, over $T_{eq}$, only if the internal temperature is at the high edge of the values expected for GJ~436b. Even if the effective temperature of the planet is just moderately enhanced by tidal heating, the latter controls the thermal structure of the deep atmosphere, where molecular abundances get vertically quenched, and thus it can affect significantly the chemical composition and in turn the emission spectrum.

\begin{figure}
\centering
\includegraphics[angle=0,width=\columnwidth]{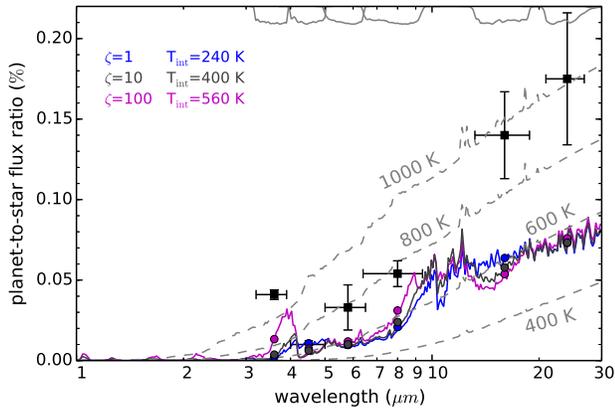}
\caption{Emission spectra calculated for GJ~436b for an intermediate case of metallicity and internal temperature of the planet ($\zeta$ = 10 and $T_{int}$ = 400 K) and two more extreme cases ($\zeta$ = 1 and $T_{int}$ = 240 K, and $\zeta$ = 100 and $T_{int}$ = 560 K). The planetary flux is shown relative to that of the star, for which a Phoenix synthetic spectrum (see Section~\ref{sec:model}) and the stellar radius given in Table~\ref{table:parameters} are adopted. Grey dashed lines indicate the planet-to-star flux ratio for planetary blackbody temperatures of 400, 600, 800, and 1000 K. The Spitzer photometric observations at 3.6, 4.5, 5.8, 8, 16, and 24 $\mu$m obtained during secondary eclipse by \cite{ste2010} are also shown as filled squares, except for the 4.5 $\mu$m value, which is an upper limit indicated by a downward triangle. Grey lines on top show the response of the Spitzer filters. Calculated flux ratios averaged over the various filters are indicated by filled circles.} \label{fig:emission}
\end{figure}

In Fig.~\ref{fig:emission} we show the emission spectrum of GJ~436b calculated for an intermediate case of metallicity and internal temperature and for two more extreme cases. The fluxes observed in the various Spitzer bands \citep{ste2010} are also shown. Some chemical information can be inferred from the relative fluxes observed in the various bands. For example, the observed emission at 3.6 $\mu$m is much higher than at 4.5 $\mu$m. Methane and carbon monoxide being efficient absorbers in the 3.6 and 4.5 $\mu$m bands, respectively, the large relative 3.6-to-4.5 $\mu$m flux ratio observed is indicative of a high CO/CH$_4$ abundance ratio in the planet's dayside. According to the calculated spectra, at high metallicities and internal temperatures (see magenta line in Fig.~\ref{fig:emission}) an emission bump (or depressed absorption) appears around 4 $\mu$m, caused by the low abundance of CH$_4$ at high $\zeta$ and $T_{int}$, which makes the relative 3.6-to-4.5 $\mu$m flux ratio to increase. Observations however show a more drastic 3.6-to-4.5 $\mu$m flux ratio, which suggests that the CO/CH$_4$ abundance ratio may be even above 1000 (the maximum value achieved in our model with $\zeta$ = 100 and $T_{int}$ = 560 K). Indeed, the retrieval analysis of \cite{mad2011} indicates that the CO/CH$_4$ ratio must be $>$10,000. We note however that the 3.6-to-4.5 $\mu$m flux ratio is less extreme in the analysis of the Spitzer data by \cite{bea2011} than in the original analysis by \cite{ste2010}, which would point to a CO/CH$_4$ abundance ratio not so extreme. A second aspect worth to note is the important emission observed in the 16 $\mu$m band, where carbon dioxide absorbs effectively, compared to the emission flux in the 5.8 and 24 $\mu$m bands, where opacity is mostly provided by water vapour. This fact suggests that the CO$_2$/H$_2$O abundance ratio must remain moderately low. According to the chemical models, high metallicities and internal temperatures favour high CO/CH$_4$ abundance ratios, and thus allow to better reproduce the observed relative 3.6-to-4.5 $\mu$m flux ratio. In fact, at metallicities as high as 1000 times over solar the agreement with observations improves significantly (see \citealt{mos2013}). However, increasing the metallicity brings also an important enhancement of the CO$_2$/H$_2$O abundance ratio and thus a worse agreement with the high observed flux at 16 $\mu$m, which indicates that there must be a limit to how rich in heavy elements can the atmosphere be.

Appart from the relative fluxes in the different bands, Fig.~\ref{fig:emission} shows that there is a poor overall agreement between calculated and observed planet-to-star flux ratios, the former being substantially lower than the latter in all Spitzer bands except at 4.5 $\mu$m. The overall shape of calculated spectra is not very different among the different cases studied. Essentially, within the ranges of metallicity and internal temperature explored ($\zeta$ =1-100 and $T_{int}$ = 100-560 K), the calculated emission spectrum of the planet at infrared wavelengths can be roughly approximated by a blackbody at a temperature of about 600 K. Spitzer observations, however, indicate that brightness temperatures are in excess of 700 K in all bands, except at 4.5 $\mu$m \citep{ste2010}. It seems clear that our models, whatever the values of $\zeta$ and $T_{int}$, underestimate the planetary emission. So, what is causing this poor agreement between calculated and observed planet-to-star flux ratios$?$

\begin{figure}
\centering
\includegraphics[angle=0,width=\columnwidth]{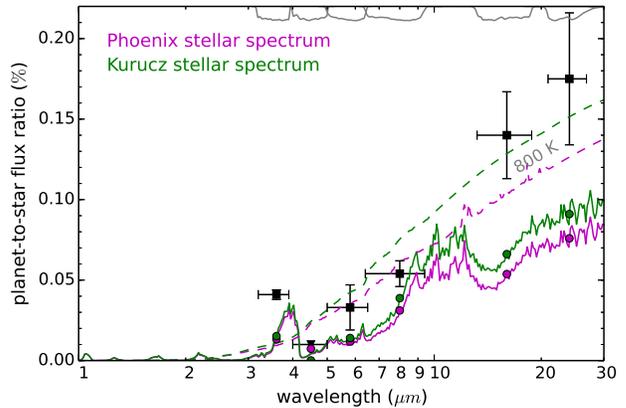}
\caption{Effect of the adopted synthetic stellar spectrum, either Phoenix (magenta) or Kurucz (green), on the planet-to-star emission flux. The planetary emission spectrum corresponds to an extreme case of metallicity and internal temperature ($\zeta$ = 100 and $T_{int}$ = 560 K). Dashed lines correspond to a planetary blackbody with a temperature of 800 K. Observed and calculated planet-to-star flux ratios in the Spitzer bands are also shown (see caption of Fig.~\ref{fig:emission}).} \label{fig:emission-phoenix-kurucz}
\end{figure}

A first important aspect to look at is the stellar spectrum adopted. Since the main outcome of secondary eclipse observations is usually given as a planet-to-star flux ratio, theoretical models aiming at interpreting these observations must rely on a synthetic emission spectrum of the star, which becomes as important as the calculated emission spectrum of the planet. In order to evaluate the influence of the adopted stellar spectrum we have considered, appart from the Phoenix synthetic spectrum described in Section~\ref{sec:model}, a Kurucz synthetic spectrum for a star with an effective temperature of 3500 K, a surface gravity of 10$^{5.0}$ cm s$^{-2}$, solar metallicity, and a microturbulence velocity of 2 km s$^{-1}$ \citep{cas2004}\footnote{See \texttt{http://wwwuser.oat.ts.astro.it/castelli/.}}, whose flux has been scaled down to get the same bolometric flux of a star with an effective temperature of 3400 K. The choice of the microturbulence velocity of 2 km s$^{-1}$ in the Kurucz spectrum is adequate for a star such as GJ 436 (F. Castelli, private communication). At infrared wavelengths, the Kurucz spectrum has a flux somewhat weaker than the Phoenix spectrum and thus using the former yields planet-to-star flux ratios higher (by $\sim$20 \% ) than using the latter(see Fig.~\ref{fig:emission-phoenix-kurucz}). Although significant, the effect of the adopted stellar spectrum cannot explain by itself the important discrepancies found between calculated and observed planet-to-star flux ratios. If we assume that the observed secondary eclipse depths are accurate within their error bars and the synthetic Phoenix and Kurucz stellar spectra provide a realistic estimate for GJ~436b, within about 20 \%, the source of the discrepancies must come from the too low planetary emission calculated.

The calculated emission spectrum of the planet consists just of a thermal emission component. A non zero planetary albedo at infrared wavelengths would add a component of reflected light to the luminosity of the planet, although its contribution to the planet-to-star flux ratio would just be $\sim$0.0002 \% for an albedo of 0.1, and the albedo of GJ~436b at infrared wavelengths is likely to be even lower based on those of Uranus and Neptune, which decrease well below 0.1 at wavelengths longer than 1 $\mu$m \citep{bin1972,wam1973}.

\begin{figure}
\centering
\includegraphics[angle=0,width=\columnwidth]{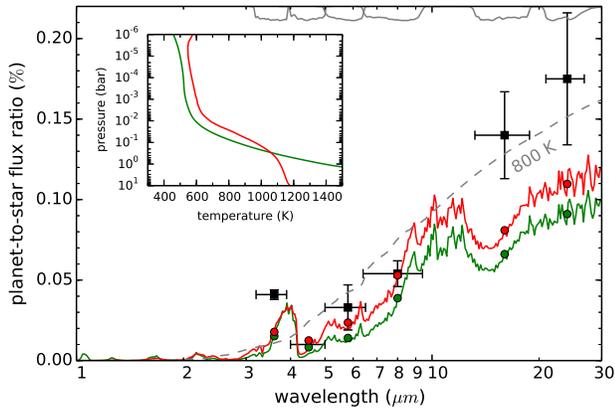}
\caption{Effect of the pressure-temperature profile on the emission spectrum. In green our calculated temperature profile and emission spectrum for an extreme case of metallicity and internal temperature of the planet ($\zeta$ = 100 and $T_{int}$ = 560 K). In red the pressure-temperature profile calculated by \cite{lew2010} for a metallicity of $\zeta$ = 50 (see text) and the emission spectrum resulting from switching to this latter temperature profile. The Kurucz synthetic spectrum described in the text is adopted for the star in both cases. The dashed line refers to a planetary blackbody temperature of 800 K. Observed and calculated planet-to-star flux ratios in the Spitzer bands are also shown (see caption of Fig.~\ref{fig:emission}).} \label{fig:emission-tklewis2010}
\end{figure}

The emission flux calculated for GJ~436b being noticeably lower than indicated by the Spitzer observations, it seems reasonable to think on a planetary atmosphere significantly warmer than calculated by our one-dimensional radiative-convective models. The thermal structure of GJ~436b's atmosphere has been investigated by \cite{lew2010} through one-dimensional radiative-convective models and general circulation models (GCMs). In the absence of dynamics and for planet-wide redistribution of the heat, these authors calculate pressure-temperature profiles (see their Figure~1) not far from ours, although with some significant differences. For super-solar metallicities, our models are warmer than theirs in the deep atmosphere, but cooler above the $\sim$1 bar pressure level, from where most of the thermal emission at infrared wavelengths arises. This is illustrated in Fig.~\ref{fig:emission-tklewis2010}, where we compare the pressure-temperature profile obtained by us for an extreme case of metallicity and internal temperature of the planet ($\zeta$ = 100 and $T_{int}$ = 560 K) with the profile calculated by \cite{lew2010} for their highest metallicity explored ($\zeta$ = 50). As shown in Fig.~\ref{fig:emission-tklewis2010}, the use of the warmer pressure-temperature profile of \cite{lew2010}, together with the choice of the Kurucz stellar spectrum, raises the calculated planet-to-star flux ratios to values in better accordance with the observed ones, although still far in the 3.6, 16, and 24 $\mu$m bands.

Although the best agreement with secondary eclipse observations is obtained in a somewhat extreme case of metallicity and internal temperature, and even forcing the atmosphere to the warmer scenario given by the temperature profile of \cite{lew2010}, this suggests that an atmosphere warmer than calculated by us may indeed help to explain the high planetary emission observed. A warming of the atmosphere can result from an effective dissipation of tidal energy, although the heating affects essentially the deep atmosphere, somewhat below the location of the photosphere, so that the impact on the emerging thermal emission is just moderate (see Fig.~\ref{fig:emission}). A more effective warming of dayside photospheric layers may result from an inefficient day-night redistribution of heat. In fact, if the incident stellar energy is distributed over just the dayside hemisphere, instead of over the whole planet, the equilibrium temperature $T_{eq}$ (assuming zero albedo) raises from 650 to 770 K, in better agreement with the brightness temperatures observed ($>$700 K). Hints on how the energy is distributed in the atmosphere of GJ~436b are provided by the GCM simulations of \cite{lew2010}, which show that the day-night temperature contrast increases, and thus the day-night heat redistribution becomes less efficient, with increasing metallicity. Thus, a combination of efficient tidal heating and inefficient day-night heat redistribution, the latter possibly driven by a high metallicity according to \cite{lew2010}, may help to explain the high dayside emission observed. High metallicities and warm temperatures favour also a high CO/CH$_4$ abundance ratio, and thus provide a better match to the observed 3.6-to-4.5 $\mu$m flux ratio. On the other hand, the high emission observed in the 16 $\mu$m band indicates that the CO$_2$/H$_2$O must remain moderately low, and therefore metallicity cannot be extremely high.

The thermal and chemical conditions needed to match the secondary eclipse observations of GJ~436b have been recently discussed by \cite{mos2013}. These authors explore a wide range of metallicities, up to $\zeta >$1000, and use pressure-temperature profiles either from one-dimensional radiative-convective models, assuming inefficient day-night redistribution of heat, or dayside averages from the GCM data of \cite{lew2010}. In either case, their temperature profiles are warmer than ours at the relevant photospheric layers. \cite{mos2013} find that the agreement with secondary eclipse observations is rather poor assuming solar metallicity (their calculated emission spectrum is quite similar to that found by us for $\zeta$ = 1) but improves significantly for metallicities as high as 1000 times solar. The improvement at such high metallicity is mainly due to, on the one hand, a 3.6-to-4.5 $\mu$m flux ratio closer to the observed one (caused by a CO/CH$_4$ abundance ratio as high as $\sim$4000), and on the other, an overall increase of the planetary emission at infrared wavelengths. In general, an increase of metallicity makes the atmosphere to be warmer (\citealt{lew2010,mos2013}; see also Fig.~\ref{fig:tk}), but also more opaque, and thus the photosphere shifts upward to upper and cooler layers, so that in principle there should be no net change in the bolometric planetary flux. That is, changes in the metallicity may bring variations of the emission flux in some bands, but variations across the whole spectrum must be counterbalanced. To this respect, we note that the higher overall planetary emission calculated by \cite{mos2013} when metallicity is increased from $\zeta$ = 1 to 1000 may be affected by a possible lack of coherency between the different radiative transfer tools and/or opacity databases used to compute temperature profiles and emission spectra at different metallicities. \cite{lew2010} have studied the effect of metallicity on the emission spectrum of GJ~436b using the temperatures from their GCM simulations and a chemical equilibrium composition. These authors find moderately low changes of the flux in the various Spitzer bands when metallicity is increased from solar to 50 times solar. We also find small changes in the emission spectrum due to variations in the metallicity from $\zeta$ = 1 to 100. In both the work of \cite{lew2010} and ours, calculated planetary fluxes in most Spitzer bands remain well below the observed values in most models, although the metallicities explored are not as high as $\zeta$ = 1000, a value at which \cite{mos2013} find higher planetary fluxes, in better agreement with observations. It is thus difficult to conclude whether at such high metallicities the thermal emission in the Spitzer bands experience a significant increase at the expense of other spectral windows, or whether the use of different radiative transfer tools and/or opacity databases by \cite{mos2013} is an issue.

Retrieval models find a relatively good agreement with the secondary eclipse observations of GJ~436b, although at the expense of temperature profiles and chemical compositions which do not follow from physical and chemical grounds \citep{ste2010,mad2011,mos2013}. Interestingly, the temperature profiles obtained by these retrieval methods are not very different from those obtained by radiative-convective models and GCM simulations at low (around solar) metallicities, although they show a greater lapse rate leading to higher photospheric temperatures, and thus higher brightness temperatures, in better agreement with observations. We stress that such temperature profiles do not arise from a plausible physical model, although they suggest the need for a warmer photosphere to reproduce the brightness temperatures derived from observations. With regard to the chemical composition, these retrieval methods indicate that a high ($\gtrsim$10,000) CO/CH$_4$ abundance ratio is needed to match the 3.6-to-4.5 $\mu$m flux ratio observed in GJ~436b's dayside.

\section{Summary} \label{sec:summary}

We have studied the influence of tidal heating and metallicity on the thermal and chemical structure of the atmosphere of the eccentric "hot Neptune" GJ~436b. Tidal heating enhances the temperatures in the bottom layers of the atmosphere and, together with metallicity, has a direct influence on the chemical equilibrium composition around the quench level, and thus affects the chemical composition in much of the atmosphere. We have explored plausible metallicities ($\zeta$ = 1-100) and internal temperatures (up to 560 K) for GJ~436b, and found that either methane or carbon monoxide can be the major carbon reservoir. The CO/CH$_4$ and N$_2$/NH$_3$ abundance ratios may take values in the range 0.05-1000 and 0.2-1000, respectively, the largest value being reached at the high edge of metallicity and internal temperature ($\zeta$ = 100 and $T_{int}$ = 560 K). Water vapour locks a large fraction of oxygen and thus remains very abundant whatever the metallicity and extent of tidal heating, while carbon dioxide experiences a very important abundance enhancement as metallicity increases.

The retrieval of information on the chemical composition of GJ~436b's atmosphere from primary transit and secondary eclipse data is complicated by the contradictory analyses of different authors on Spitzer data. In general, the agreement of calculated transmission and emission spectra with available observations is somewhat poor, although some conclusions can be extracted. In the case of transmission spectra, the relative variation of the transit depth in the 3.6 and 4.5 $\mu$m bands measured by \cite{bea2011}, which suggests a methane-rich atmosphere, cannot be reproduced by our model with the lowest CO/CH$_4$ abundance ratio, while that measured by \cite{knu2011}, which suggests a methane-poor atmosphere, may be consistent with models with a high metallicity and efficient tidal heating, in which the CO/CH$_4$ abundance ratio is high. As concerns emission spectra, the 3.6-to-4.5 $\mu$m flux ratio measured by \cite{ste2010} suggests a CO/CH$_4$ abundance ratio higher than 1000 (the largest value calculated by us at the high edge of metallicity and internal temperature), which may point to metallicities above 100 times solar, while the observational analysis of \cite{bea2011} is less demanding in terms of high CO/CH$_4$ abundance ratios. On the other hand, a very high metallicity brings a great enhancement of the CO$_2$/H$_2$O abundance ratio, and thus a worse agreement with the high planetary flux observed at 16 $\mu$m. Overall, the calculated planetary emission is lower than observed during secondary eclipse, which points to a rather warm dayside atmosphere, probably due to a combination of efficient tidal heating and inefficient day-night heat redistribution, the latter possibly driven by high metallicities, according to \cite{lew2010}.

Although none of the atmospheric models of GJ~436b based on solid physical and chemical grounds provide a fully satisfactory agreement with available observational data (see also \citealt{mos2013}), observations seem to favour models with a high metallicity and efficient tidal heating. It must however be recognised that the chemical composition of GJ~436b's atmosphere remains puzzling to date, and future observations with telescopes such as the James Webb Space Telescope and EChO are urgently needed to draw a definitive picture about the main atmospheric constituents of this peculiar "hot Neptune".

\acknowledgements

We thank our referee for a constructive report that helped to improve this paper. We are grateful to Vincent Eymet and Philip von Paris for their very useful input regarding line-by-line radiative transfer calculations and Nikku Madhusudhan for useful correspondance concerning the emission spectrum of GJ~436b. M.A. and F.S. acknowledge support from the European Research Council (ERC Grant 209622: E$_3$ARTHs). O. V. acknowledges support from the KU Leuven IDO project IDO/10/2013 and from the FWO Postdoctoral Fellowship programme. Computer time for this study was provided by the computing facilities MCIA (M\'esocentre de Calcul Intensif Aquitain) of the Universit\'e de Bordeaux and of the Universit\'e de Pau et des Pays de l'Adour.


\begin{thebibliography}{}

\bibitem[Ag\'undez et al.(2012)]{agu2012} Ag\'undez, M., Venot, O., Iro, N., et al. 2012, \aap, 548, A73
\bibitem[Ag\'undez et al.(2013)]{agu2013} Ag\'undez, M., Parmentier, V. Venot, O., et al. 2013, \aap, submitted
\bibitem[Alonso et al.(2008)]{alo2008} Alonso, R., Barbieri, M., Rabus, M., et al. 2008, \aap, 487, L5
\bibitem[Asplund et al.(2009)]{asp2009} Asplund, M., Grevesse, N., Sauval, A. J., \& Scott, P. 2009, \araa, 47, 481
\bibitem[Ballard et al.(2010)]{bal2010} Ballard, S., Christiansen, J. L., Charbonneau, D., et al. 2010, \apj, 716, 1047
\bibitem[Beaulieu et al.(2011)]{bea2011} Beaulieu, J.-P., Tinetti, G., Kipping, D. M., et al. 2011, \apj, 731, 16
\bibitem[Beust et al.(2012)]{beu2012} Beust, H., Bonfils, X., Montagnier, G., Delfosse, X., \& Forveille, T. 2012, \aap, 545, A88
\bibitem[Binder \& McCarthy(1972)]{bin1972} Binder, A. B. \& McCarthy, D. W. 1972, \apj, 171, L1
\bibitem[Borysow et al.(1989)]{bor1989} Borysow, A., Frommhold, L., \& Moraldi, M. 1989, \apj, 336, 495
\bibitem[Borysow \& Frommhold(1989)]{borfro1989} Borysow, A. \& Frommhold, L. 1989, \apj, 341, 549
\bibitem[Borysow et al.(1997)]{bor1997} Borysow, A., J{\o}rgensen, U. G., \& Zheng, C. 1997, \aap, 324, 185
\bibitem[Borysow et al.(2001)]{bor2001} Borysow, A., J{\o}rgensen, U. G., \& Fu, Y. 2001, \jqsrt, 68, 235
\bibitem[Borysow(2002)]{bor2002} Borysow, A. 2002, \aap, 390, 779
\bibitem[Butler et al.(2004)]{but2004} Butler, R. P., Vogt, S. S., Marcy, G. W., et al. 2004, \apj, 617, 580
\bibitem[C\'aceres et al.(2009)]{cac2009} C\'aceres, C., Ivanov, V. D., Minniti, D., et al. 2009, \aap, 507, 481
\bibitem[Castelli \& Kurucz(2004)]{cas2004} Castelli, F. \& Kurucz, R. L. 2004, \texttt{arXiv:astro-ph/0405087}
\bibitem[Colombo(1966)]{col1966} Colombo, G. 1966, \aj, 71, 891
\bibitem[Deming et al.(2007)]{demi2007} Deming, D., Harrington, J., Laughlin, G., et al. 2007, \apj, 667, L199
\bibitem[Demory et al.(2007)]{demo2007} Demory, B.-O., Gillon, M., Barman, T., et al. 2007, \aap, 475, 1125
\bibitem[Elkins-Tanton \& Seager(2008)]{elk2008} Elkins-Tanton, L. T. \& Seager, S. 2008, \apj, 685, 1237
\bibitem[Eymet et al.(2009)]{eym2009} Eymet, V., Fournier, R., Dufresne, J.-L., et al. 2009, \jgr~Planets, 114, 11008
\bibitem[France et al.(2013)]{fra2013} France, K., Froning, C. S., Linsky, J. L., et al. 2013, \apj, 763, 149
\bibitem[Gardner et al.(2006)]{gar2006} Gardner, J. P., Mather, J. C., Clampin, M., et al. 2006, \ssr, 123, 485
\bibitem[Gillon et al.(2007)]{gil2007} Gillon, M., Pont, F., Demory, B.-O., et al. 2007, \aap, 472, L13
\bibitem[Goldreich \& Soter(1966)]{gol1966} Goldreich, P. \& Soter, S. 1966, \icarus, 5, 375
\bibitem[Hauschildt et al.(1999)]{hau1999} Hauschildt, P. H., Allard, F., \& Baron, E. 1999, \apj, 512, 377
\bibitem[Hersant et al.(2004)]{her2004} Hersant, F., Gautier, D., \& Lunine, J. I. 2004, \pss, 52, 623
\bibitem[Hut(1981)]{hut1981} Hut, P. 1981, \aap, 99, 126
\bibitem[Iro et al.(2005)]{iro2005} Iro, N., B\'ezard, B., \& Guillot, T. 2005, \aap, 436, 719
\bibitem[Iro \& Deming(2010)]{iro2010} Iro, N. \& Deming, L. D. 2010, \apj, 712, 218
\bibitem[Knutson et al.(2011)]{knu2011} Knutson, H. A., Madhusudhan, N., Cowan, N. B., et al. 2011, \apj, 735, 27
\bibitem[Leconte et al.(2010)]{lec2010} Leconte, J., Chabrier, G., Baraffe, I., \& Levrard, B. 2010, \aap, 516, A64
\bibitem[Levrard et al.(2007)]{lev2007} Levrard, B., Correia, A. C. M., Chabrier, G., et al. 2007, \aap, 462, L5
\bibitem[Lewis et al.(2010)]{lew2010} Lewis, N. K., Showman, A. P., Fortney, J. J., et al. 2010, \apj, 720, 344
\bibitem[Line et al.(2011)]{lin2011} Line, M. R., Vasisht, G., Chen, P., et al. 2011, \apj, 738, 32
\bibitem[Madhusudhan \& Seager(2011)]{mad2011} Madhusudhan, N. \& Seager, S. 2011, \apj, 729, 41
\bibitem[Maness et al.(2007)]{man2007} Maness, H. L., Marcy, G. W., Ford, E. B., et al. 2007, \pasp, 119, 90
\bibitem[Mardling(2008)]{mar2008} Mardling, R. A. 2008, \texttt{arXiv:0805.1928}
\bibitem[Moses et al.(2013)]{mos2013} Moses, J. I., Line, M. R., Visscher, C., et al. 2013, \apj, 777, 34
\bibitem[Nettelmann et al.(2010)]{net2010} Nettelmann, N., Kramm, U., Redmer, R., \& Neuh\"auser, R. 2010, \aap, 523, A26
\bibitem[\"Onehag et al.(2012)]{one2012} \"Onehag, A., Heiter, U., Gustafsson, B., et al. 2012, \aap, 542, A33
\bibitem[Parmentier et al.(2013)]{par2013} Parmentier, V., Showman, A. P., \& Lian, Y. 2013, \aap, 558, A91
\bibitem[Peale(1969)]{pea1969} Peale, S. J. 1969, \aj, 74, 483
\bibitem[Pont et al.(2009)]{pon2009} Pont, F., Gilliland, R. L., Knutson, H., Holman, M., \& Charbonneau, D. 2009, \mnras, 393, L6
\bibitem[Rothman et al.(2009)]{rot2009} Rothman, L. S., Gordon, I. E., Barbe, A., et al. 2009, J. Quant. Spec. Radiat. Transf., 110, 533
\bibitem[Rothman et al.(2010)]{rot2010} Rothman, L. S., Gordon, I. E., Barber, R. J., et al. 2010, J. Quant. Spec. Radiat. Transf., 111, 2139
\bibitem[Shabram et al.(2011)]{sha2011} Shabram, M., Fortney, J. J., Greene, T. P., \& Freedman, R. S. 2011, \apj, 727, 65
\bibitem[Southworth(2010)]{sou2010} Southworth, J. 2010, \mnras, 408, 1689
\bibitem[Stevenson et al.(2010)]{ste2010} Stevenson, K. B., Harrington, J., Nymeyer, S., et al. 2010, \nature, 464, 1161
\bibitem[Thuillier et al.(2004)]{thu2004} Thuillier, G., Floyd, L., Woods, T. N., et al. 2004, \asr, 34, 256
\bibitem[Tinetti et al.(2012)]{tin2012} Tinetti, G., Beaulieu, J. P., Henning, T., et al. 2012, \expa, 34, 311
\bibitem[Venot et al.(2012)]{ven2012} Venot, O., H\'ebrard, E., Ag\'undez, M., et al. 2012, \aap, 546, A43
\bibitem[Venot et al.(2013a)]{ven2013a} Venot, O., Fray, N., B\'enilan, Y., et al. 2013a, \aap, 551, A131
\bibitem[Venot et al.(2013b)]{ven2013b} Venot, O., Ag\'undez, M., Selsis, F., et al. 2013b, \aap, submitted
\bibitem[von Braun et al.(2012)]{von2012} von Braun, K., Boyajian, T. S., Kane, S. R., et al. 2012, \apj, 753, 171
\bibitem[Wamsteker(1973)]{wam1973} Wamsteker, W. 1973, \apj, 184, 1007
\bibitem[Zahnle et al.(2009)]{zah2009} Zahnle, K., Marley, M. S., Freedman, R. S., Lodders, K., \& Fortney, J. J. 2009, \apj, 701, L20
\bibitem[Zhang \& Hamilton(2008)]{zha2008} Zhang, K. \& Hamilton, D. P. 2008, \icarus, 193, 267

\end{thebibliography}
\end{document}